\documentclass[12pt,a4paper,final]{iopart}

\usepackage{iopams}
\usepackage{graphicx}
\usepackage[breaklinks=true,colorlinks=true,linkcolor=blue,urlcolor=blue,citecolor=blue]{hyperref}

\begin{document}

\title
{One-parameter supersymmetric Hamiltonians in momentum space}

\author{H. C. Rosu}
\address{IPICYT, Instituto Potosino de Investigacion Cientifica y Tecnologica,\\
Camino a la presa San Jos\'e 2055, Col. Lomas 4a Secci\'on, 78216 San Luis Potos\'{\i}, S.L.P., Mexico}
\ead{hcr@ipicyt.edu.mx}

\author{S.C. Mancas}
\address{Department of Mathematics, Embry-Riddle Aeronautical University, Daytona Beach, FL 32114-3900, USA}
\ead{mancass@erau.edu}

\author{P. Chen}
\address{Leung Center for Cosmology and Particle Astrophysics (LeCosPA) and Department of Physics, National Taiwan University, Taipei 10617, Taiwan}
\ead{pisinchen@phys.ntu.edu.tw}


\begin{abstract}
Recent results on the one-parameter supersymmetric deformation in momentum space by
Curtright and Zachos (2014 \textit{J. Phys. A: Math. and Theor.} {\bf 47} 145201) are presented in a more general framework following our own papers. We extend the analysis of Curtright and Zachos by including the supersymmetric partner one-parameter deformation.
\end{abstract}

\vspace{2pc}

\vspace{2pc}

Supersymmetric quantum mechanics \cite{B1,B2} is an algebraic operator method which has been used to generate new exactly-solvable Schr\"odinger spectral problems. It is predominantly performed in configuration space because algebraically one does not expect differences when one changes the representation. The basic commutators, such as $[x,p]$ which turns a wavefunction in momentum space $\phi(p)$ into $i\phi(p)$, act exactly as in the configuration space, and therefore any more complicated algebraic structure is preserved in form. As Weigold \cite{w82} expressed many years ago the coordinate representation of quantum mechanics prevails over the momentum representation because ``we are used to think in position space but not in the momentum space", although in all the important experimental methods, such as electron-momentum spectroscopy \cite{vos97} and Compton scattering \cite{dumond33}, the momentum distribution of the electrons is the primarily measured quantity.

In a recent paper by Curtright and Zachos (CZ)~\cite{ref1}, the deformed one-parameter supersymmetry based on the general Riccati solution is performed in the momentum space of quantum mechanics and not in the usual coordinate representation. The motivation can be found in their focus on the interesting topic of branched (multi-valued) Hamiltonians which has been revived by Shapere and Wilczek \cite{sw1,sw2,w3} and requires to go to the momentum space. The reason behind this is that in this way the strange problem of studying multiple-valued Hamiltonians is turned into standard quantum-mechanical problems with special attention to the probability flow at the boundaries. Essentially, CZ discuss the quantization in momentum space for a classical Hamiltonian with two kinetic energy branches and a square power potential, where one is led to a supersymmetric pair of quantum Hamiltonians in momentum space. However, they also employ the one-parameter supersymmetric method in which the pair of supersymmetric partner potentials is turned into a one-parameter family of potentials, which is equivalent to a family of infinite number of branches labeled by the values of the parameter.

The purpose of this paper is to present the CZ case from the point of view of our recent papers \cite{rmc1,rmc2} providing in this way a more detailed analysis of this interesting case. Moreover, our analysis yield up features of the method which go beyond those in \cite{ref1}. In particular, we will show that a family of deformed supersymmetric branches different from the CZ one is possible.

\begin{figure}[x!] 
\begin{center}
\includegraphics{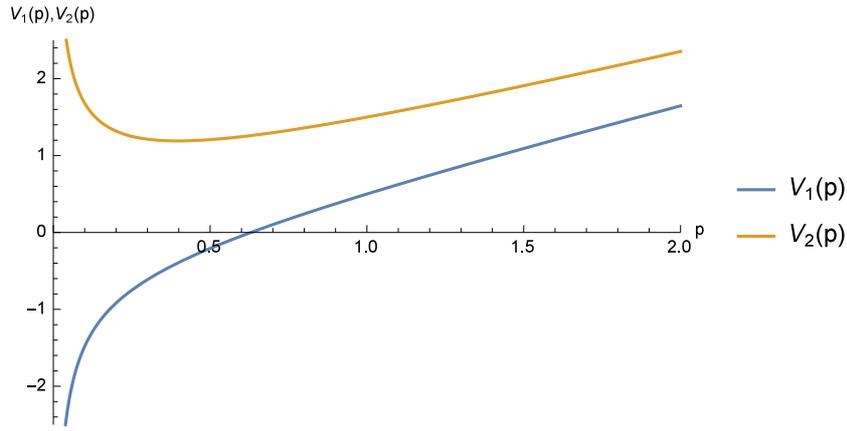}
\caption{\textsl{
The supersymmetric partner potentials $V_1(p)$ (blue) and $V_2(p)$ (orange).}}
\label{F1}
\end{center}
\end{figure}

Curtright and Zachos discuss the following supersymmetric pair
\begin{equation}
H_{\mp}(p)=-\frac{d^2}{dp^2}+p\mp\frac{1}{2\sqrt{p}}~
\end{equation}
forming the two branches of the corresponding supersymmetric Hamiltonian that can be represented as a $2\times 2$ diagonal matrix with the two partner Hamiltonians set on the diagonal positions.
The usual factorization of these Hamiltonians is given by
\begin{equation}
H_{\mp}(p)=\left(\frac{d}{dp}\mp W_0(p)\right)\left(-\frac{d}{dp}\mp W_0(p)\right)
\end{equation}
where the particular Riccati solution $W_0(p)=\sqrt{p}$ gets involved.
Alternatively, one can say that the pair of supersymmetric partner potentials are given by the following pair of Riccati equations
\begin{eqnarray}
-&W'+W^2=V_1 \label{r1}\\
&W'+W^2=V_2 \label{r2}~,
\end{eqnarray}
where the prime stands for the derivative with respect to $p$. These equations correspond to $H_-$ and $H_+$, respectively.
If one uses the particular solution $W_0(p)$, one gets the potentials
\begin{eqnarray}
&V_1(p)=p-\frac{1}{2\sqrt{p}}\label{v1}\\
&V_2(p)=p+\frac{1}{2\sqrt{p}}~.\label{v2}
\end{eqnarray}
These partner potentials are displayed in figure~\ref{F1}.
However, since the works of Mielnik \cite{miel}, Fern\'andez \cite{fern}, and Nieto \cite{n84}, it is known that one can construct one-parameter families of potentials if one employs the general Riccati solution of either Eq.~(\ref{r1}) or Eq.~(\ref{r2}). In our recent papers on the deformed one-parameter supersymmetric potentials \cite{rmc1,rmc2}, we used the Riccati equation (\ref{r2}) as the starting point for this construction, while in section 5 of their paper, Curtright and Zachos use the first Riccati equation (\ref{r1}). For the sake of completeness, we discuss here both deformations on an equal footing using some general results presented in \cite{rmc1,rmc2}.

\medskip

Substituting the Bernoulli ansatz for the general Riccati solution $W=W_0+1/u$ in the two Riccati equations, one gets the following linear equations
\begin{eqnarray}
& u'+2W_0u+1=0\label{l1}\\
-&u'+2W_0u+1=0\label{l2}~.
\end{eqnarray}
The integrating factors of these equations are $\mu_1(p)=e^{2\int^p W_0dp}$ and $\mu_2(p)=e^{-2\int^p W_0dp}$, respectively, which for the CZ case are explicitly
\begin{equation}
\mu_1(p)=\exp\left(\frac{4}{3}p^{\frac{3}{2}}\right)~, \qquad \mu_2(p)=\exp\left(-\frac{4}{3}p^{\frac{3}{2}}\right)~.
\end{equation}
The square roots of these integrating factors are the zero modes (ground states) of $H_+$ and $H_-$, i.e.,
\begin{equation}
\tilde{\Psi}_0\equiv\sqrt{\mu_1}=\exp\left(\frac{2}{3}p^{\frac{3}{2}}\right)~ \quad {\rm and} \quad
\Psi_0\equiv\sqrt{\mu_2}=\exp\left(-\frac{2}{3}p^{\frac{3}{2}}\right)~,
\end{equation}
respectively \cite{rmc1}.
Then, one can obtain the general Riccati solutions in the form
\begin{equation}
W_{1\gamma}=W_0+\frac{\mu_1}{\gamma-\int^p \mu_1dp}~, \qquad W_{2\gamma}=W_0+\frac{\mu_2}{\gamma+\int^p \mu_2dp}~,
\end{equation}
which can be also written as
\begin{equation}
W_{1\gamma}=W_0-\frac{d}{dp}\ln\left|\gamma-\int^p \mu_1dp\right|~, \quad W_{2\gamma}=W_0+\frac{d}{dp}\ln\left|\gamma+\int^p \mu_2dp\right|~,
\end{equation}
where $\gamma$ is the integration constant parameter, which is the inverse of the parameter used by Curtright and Zachos and can be used as a deformation parameter in the supersymmetric construction.
It is easy to show then using $V_{1\gamma}=V_2-2\frac{d W_{2\gamma}}{dp}$ and $V_1=V_2-2\frac{dW_0}{dp}$ that the first parametric family of supersymmetric potentials has the general expression 
\begin{eqnarray}
&V_{1\gamma}(p)=V_1(p)-2\frac{d^2}{dp^2}\ln\left |\gamma+\gamma_2(p)\right|~, \label{13}\\
&\gamma_2(p)=\int_0^p \mu_2(p)dp=-\frac 2 3  p E_{\frac 1 3}\left(\frac {4p^{\frac 3 2 }}{3}\right)+\frac{\Gamma\left (\frac 2 3\right )}{6 ^{\frac 1 3}}~,\label{14}
\end{eqnarray}
whereas for the second family one should use $V_{2\gamma}=V_1+2\frac{dW_{1\gamma}}{dp}$ and $V_1=V_2-2\frac{dW_0}{dp}$, which leads to
\begin{eqnarray}
&V_{2\gamma}(p)=V_2(p)-2\frac{d^2}{dp^2}\ln\left|\gamma-\gamma_1(p)\right|~, \label{15}\\
&\gamma_1(p)=\int_0^p \mu_1(p)dp=-\frac 2 3  p E_{\frac 1 3}\left(-\frac {4p^{\frac 3 2 }}{3}\right)+
\frac{\Gamma\left (\frac 2 3\right )}{6 ^{\frac 1 3}}~.\label{16}
\end{eqnarray}
In (\ref{14}) and (\ref{16}), $E_q(z)$ is the exponential integral function defined as
\begin{equation}
E_q(z)=\int_1^{\infty}\frac{\exp {(-zt})}{t^q}dt
\end{equation}
which is the upper incomplete Gamma function using the rescaling $zt \rightarrow t$
\begin{equation}
E_q(z)=z^{q-1}\int_z^{\infty}\frac{\exp {(-t})}{t^q}dt=z^{q-1}\Gamma (1-q,z)~.
\end{equation}
In our case, since $q=1/3$ we obtain the simple relationship
\begin{equation}
E_{\frac 1 3}(z)=\frac{\Gamma \left (\frac 2 3,z\right)}{\sqrt[3]{z^2}}
\end{equation}
which can be also used in (\ref{14}) and (\ref{16}) if one wants to express the results only in terms of Gamma functions.
One can also define
\begin{equation}
\Delta V_1\equiv V_{1\gamma}-V_1=-2\frac{d^2}{dp^2}\ln\left |\gamma+\gamma_2(p)\right|
 \end{equation}
and
\begin{equation}
\Delta V_2\equiv V_{2\gamma}-V_2=-2\frac{d^2}{dp^2}\ln\left |\gamma-\gamma_1(p)\right|
\end{equation}
as the one-parameter deformations of the supersymmetric partner potentials.
Moreover, the non-normalized deformed zero-modes can be written as follows
\begin{equation}
\Psi_{0\gamma}(p)=\frac{\sqrt{\mu_{2}(p)}}{\gamma+\gamma_2(p)}~, \qquad \tilde{\Psi}_{0\gamma}(p)=\frac{\sqrt{\mu_{1}(p)}}{\gamma-\gamma_1(p)}~,
\end{equation}
upon noting that the undeformed zero modes are connected to the particular Riccati solution as $\ln \Psi_0=-W_0$ and $\ln \tilde{\Psi}_0=W_0$, and then in the deformed case the relationships are $\ln \Psi_{0\gamma}=-W_{2\gamma}$ and $\ln \tilde{\Psi}_{0\gamma}=W_{1\gamma}$, according to the corresponding second logarithmic derivatives that provide the deformation of the potentials. Furthermore, an iteration process of these supersymmetric deformations is possible leading to multiple-parameter deformed potentials and zero modes \cite{r2000}.

\begin{figure}[x!] 
\begin{center}
\resizebox*{0.30\textheight}{!}{
{\includegraphics{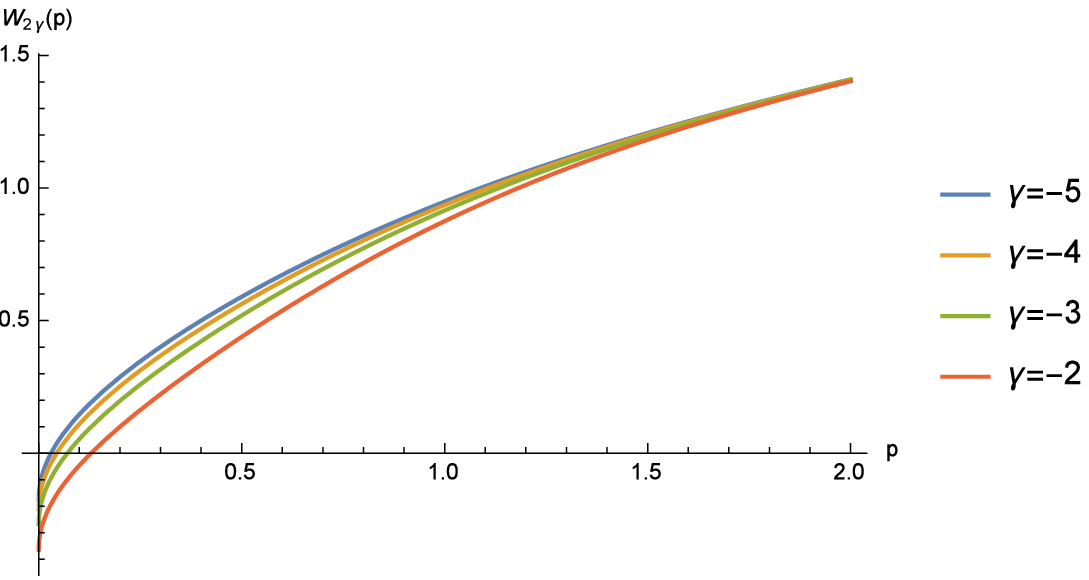}}}
\resizebox*{0.30\textheight}{!}{
{\includegraphics{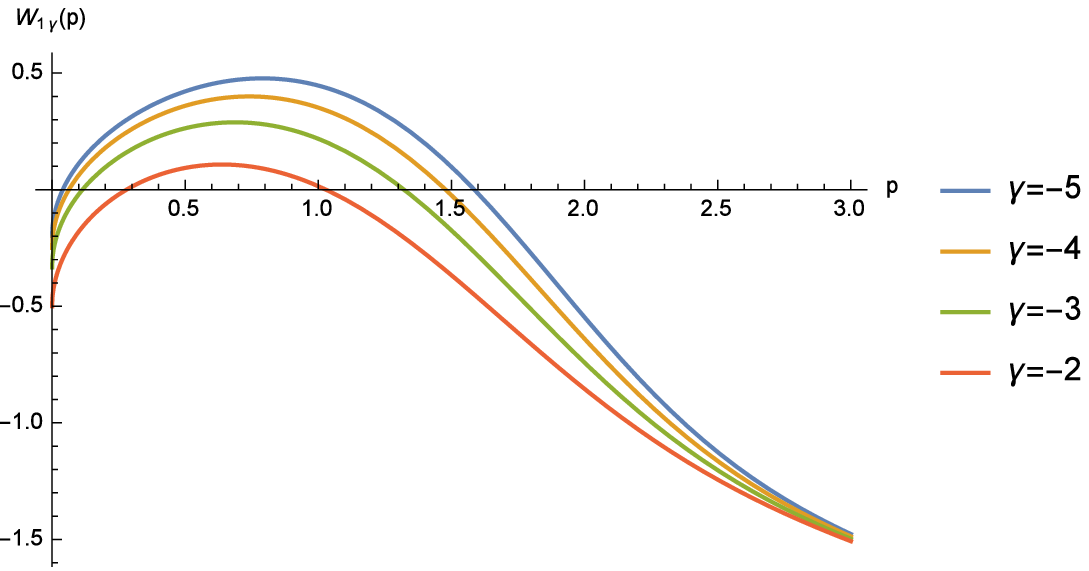}}}
\resizebox*{0.30\textheight}{!}{
{\includegraphics{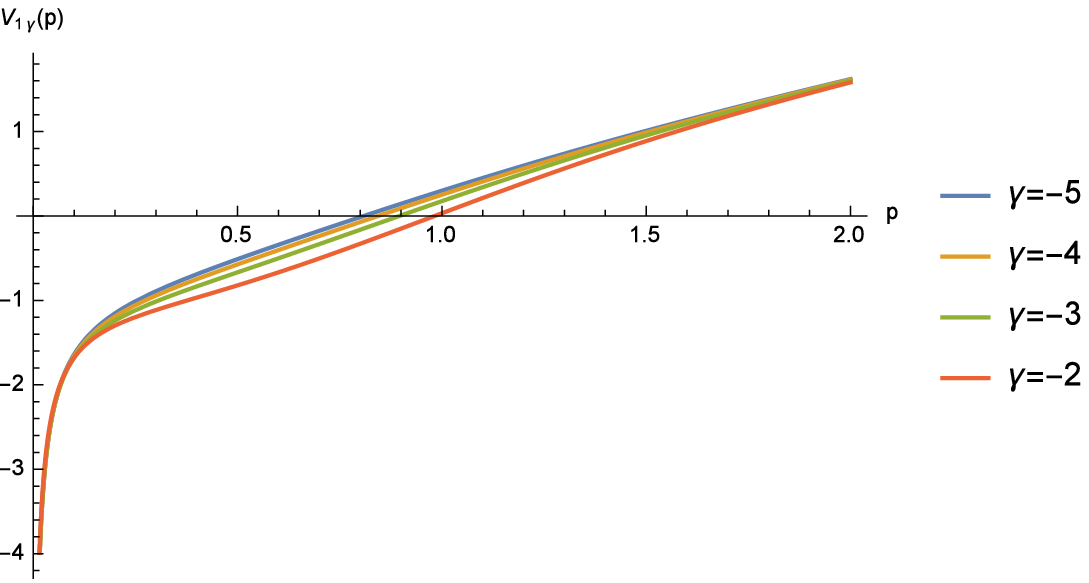}}}
\resizebox*{0.30\textheight}{!}{
{\includegraphics{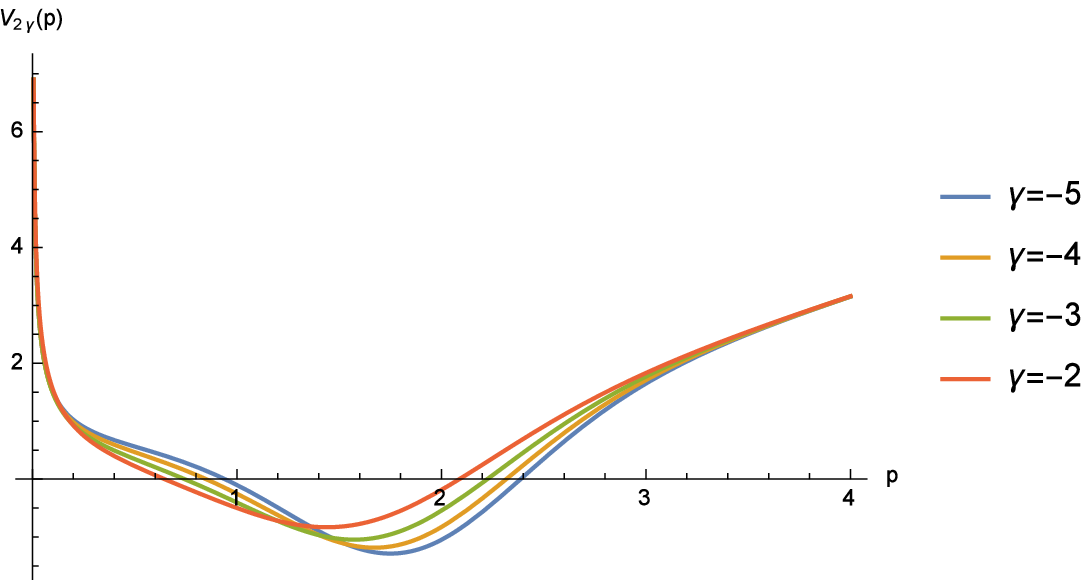}}}
\resizebox*{0.30\textheight}{!}{
{\includegraphics{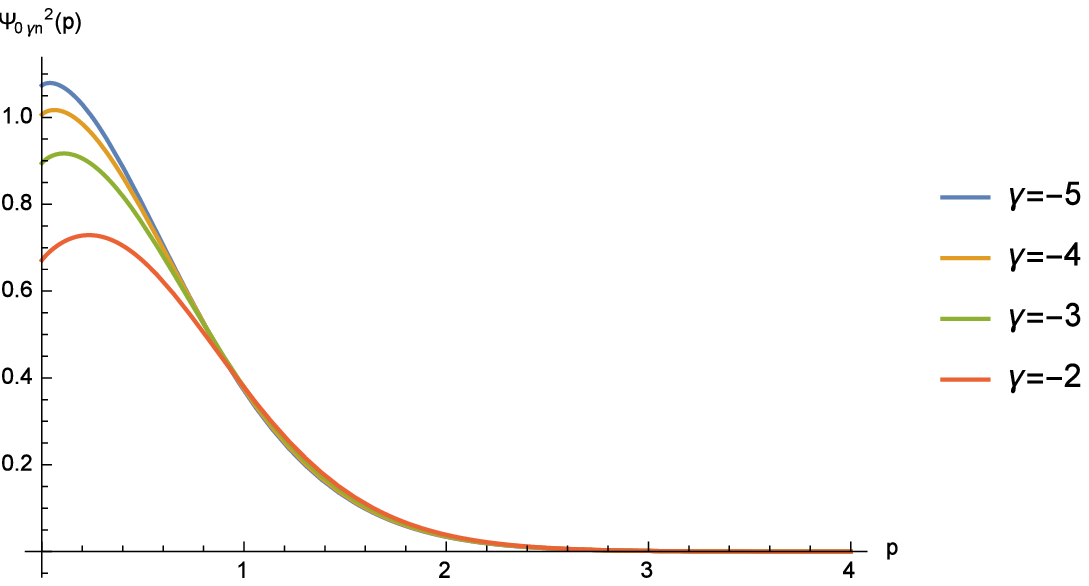}}}
\resizebox*{0.30\textheight}{!}{
{\includegraphics{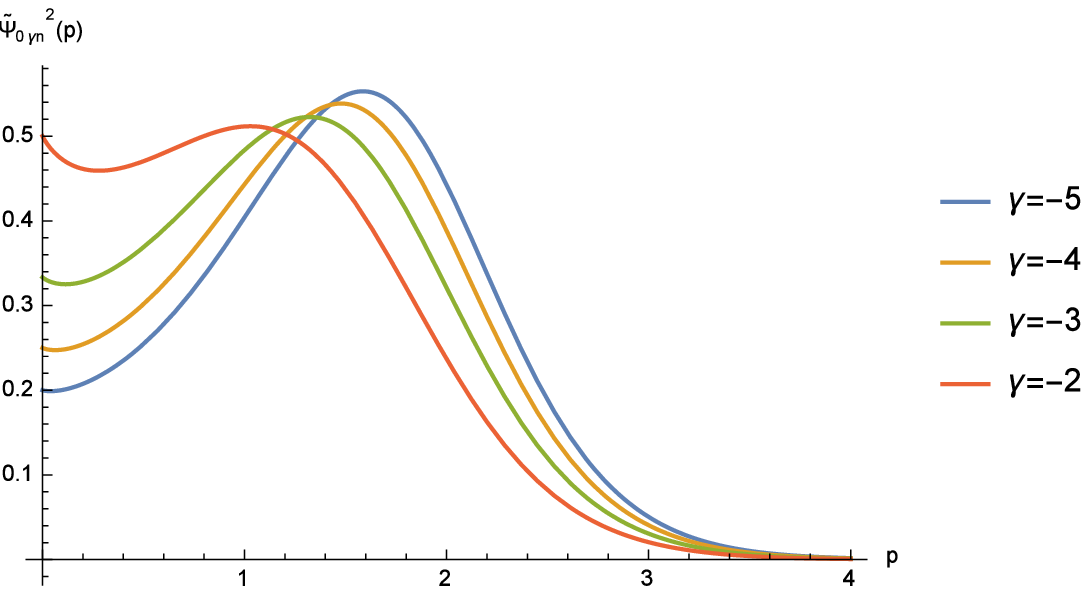}}}
\caption{\textsl{
Left panels: General Riccati solutions $W_{2\gamma}(p)$, the related deformed potentials $V_{1\gamma}(p)$, and the deformed normalized zero modes $\Psi_{0\gamma n }(p)$ for a set of four values of the deformation parameter.
Right panels: General Riccati solution $W_{1\gamma}(p)$, the related deformed potentials $V_{2\gamma}(p)$, and the deformed normalized zero modes $\tilde{\Psi}_{0\gamma n }(p)$ for the same set of values of the deformation parameter.
}}
\label{F2}
\end{center}
\end{figure}

We end up with the normalization issue of the deformed zero modes. In the case of $\Psi_{0\gamma}$, this issue is well settled, see \cite{rmc2}. One should consider first a normalized undeformed zero mode, i.e.,
\begin{equation}
\Psi_{0n}=\sqrt{\frac{\mu_2}{\gamma_2(\infty)}}
\end{equation}
and then one can show that the normalized deformed zero modes are given by
\begin{equation}
\Psi_{0\gamma n}=\sqrt{\frac{\gamma(\gamma+1)}{\gamma_2(\infty)}}\frac{\sqrt{\mu_2}}{\gamma+\gamma_2}~.
\end{equation}
From the latter expression, one can see that $\gamma$ should not be in the interval [-1,0] in standard quantum mechanical applications.
On the other hand, for $\tilde{\Psi}_{0\gamma}$, although the undeformed zero mode cannot be normalized the deformed zero modes are bounded functions at infinity. The normalization can be performed by the standard formula
\begin{equation}
\tilde{\Psi}_{0\gamma n}= \frac{1}{\sqrt{\displaystyle{} \int_0^\infty\frac{\mu_{1}}{(\gamma-\gamma_1)^2}}}\frac{\sqrt{\mu_{1}}}{\gamma-\gamma_1}~.
\end{equation}
In particular, for the case studied by Curtright and Zachos, the normalized deformed zero modes are
\begin{eqnarray}
\Psi_{0\gamma n}(p)&= \sqrt{\gamma(\gamma+1)}\sqrt{\frac{6^{\frac 1 3 }}{\Gamma\left(\frac{2}{3}\right)}}\frac{e^{-\frac{2}{3}p^{\frac{3}{2}}}}{\gamma-\frac 2 3  p E_{\frac 1 3}\left(\frac {4p^{\frac 3 2 }}{3}\right)+\frac{\Gamma\left (\frac 2 3\right )}{6 ^{\frac 1 3}}}~,\\
 \tilde{\Psi}_{0\gamma n}(p)&=\frac{1}{\sqrt{ 
 \int_0^{\infty}\frac{e^{\frac{4p^{\frac 3 2}}{3 }}dp}{\left(\gamma+\frac{2p}{3} E_{\frac 1 3}\left(-\frac {4p^{\frac 3 2 }}{3}\right)-\frac{\Gamma\left (\frac 2 3\right )}{6 ^{\frac 1 3}}\right)^2}}}
 \frac{e^{\frac{2p^{\frac 3 2}}{3}}}{\gamma+\frac{2p}{3} E_{\frac 1 3}\left(-\frac {4p^{\frac 3 2 }}{3}\right)-\frac{\Gamma\left (\frac 2 3\right )}{6 ^{\frac 1 3}}}~.
\end{eqnarray}
Plots of the general Riccati solutions and the resulting deformed potentials and zero modes for some values of the deformation parameter are presented in figure~\ref{F2}. The parameters are chosen such as to avoid the generation of singularities \cite{rmc1,rmc2}. Since $\gamma_2$ belongs to [0, $\gamma_2(\infty)$]=[0,0.7452], then $\gamma$ should not be in the interval [-0.7452,0] in the case of the parametric deformation of $V_1$. This interval is however enclosed in the one already prohibited through the normalization condition. On the other hand, because $\gamma_1>0$ for any $p$, then $\gamma$ should be strictly negative to generate continuous deformations of $V_2$.

\begin{figure}[x!] 
\begin{center}
\includegraphics{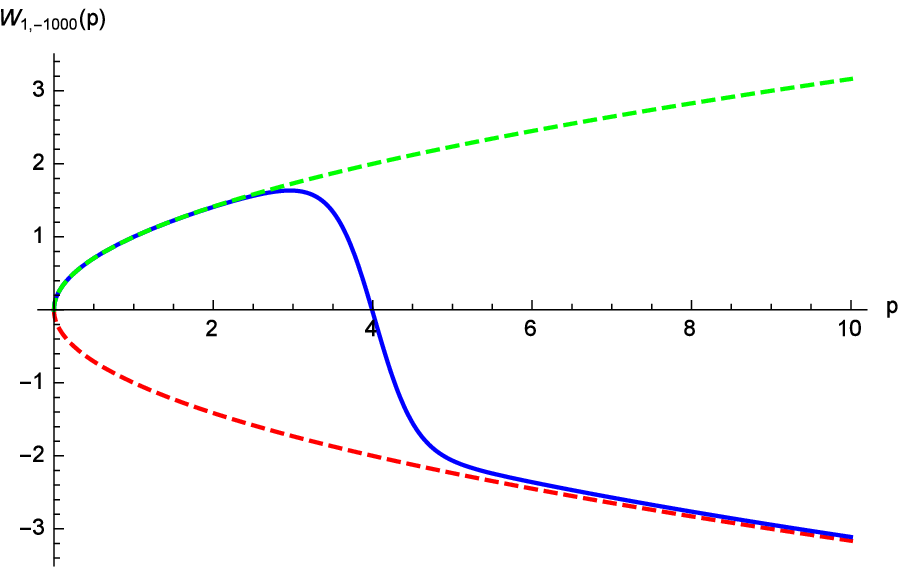}
\caption{\textsl{
The general Riccati solution $W_{1\gamma}(p)$ (blue) for $\gamma=-1000$ displaying the bending feature.}}
\label{F3}
\end{center}
\end{figure}

Regarding the large $p$ asymptotics of the two classes of deformed potentials in the CZ case, both go to the linear potential $\sim p$. On the other hand, the general Riccati solutions have the behaviour $W_{1\gamma} \rightarrow -\sqrt{p}$ and $W_{2\gamma} \rightarrow \sqrt{p}$, respectively. Moreover, both parametric Riccati solutions have initial values $1/\gamma$ but they rapidly tend for any allowed $\gamma$ to the parabolic branches $\pm \sqrt{p}$. In addition, we notice that for large $\gamma$ the functions $W_{1\gamma}$ follow at low momenta the $\sqrt{p}$ branch but at some critical $p$ they bend towards the $-\sqrt{p}$ branch. A typical example is shown in figure \ref{F3}. The reason is that at high $p$ the parametric part of $W_{1\gamma}$ tends to $-2\sqrt{p}$ forcing the bending towards the $-\sqrt{p}$ branch while in the case of $W_{2\gamma}$ the parametric part tends to zero. Concerning the deformed zero modes, in the large deformation parameter asymptotics they turn into the corresponding undeformed zero modes.

\medskip

In conclusion, we have discussed here in a more general setting the recent example of Curtright and Zachos of one-parameter deformed supersymmetry in momentum space presenting a second deformation that can be performed, the one in the left panels of figure~\ref{F2}, and pointing out the bending asymptotic feature of the general Riccati solutions $W_{1\gamma}$. Regarding the relevance of these deformations, one should notice that the deformed wavefunctions are subject to Robin boundary conditions at the origin, as it is easily established for the deformed zero modes \cite{ref1}. As such, the deformed systems might be a class of more flexible branches as to the undeformed partner systems. The latter, despite being linked through the supercharge operators, are completely separated by the boundary conditions at the origin, Dirichlet for one and Neumann for the other.

%
%
%
%

%


\section*{References}
\begin{thebibliography}{10}

\bibitem{B1} Bagchi B 2001 {\it Supersymmetry in Quantum and Classical Mechanics} (Boca Raton: Chapman and Hall/CRC)

\bibitem{B2} Cooper F, Khare A, Sukhatme U 2001  {\it Supersymmetry in Quantum Mechanics} (Singapore: World Scientific)

\bibitem{w82} Weigold E 1982 Electron coincidence spectroscopy: An introduction to momentum space chemistry \textit{Aust. J. Phys.} {\bf 35} 571-591

\bibitem{vos97} Vos M, McCarthy I 1997 Measuring orbitals and bonding in atoms, molecules, and solids \textit{Am. J. Phys.} {\bf 65} 544-553

\bibitem{dumond33} DuMond J W M 1933 The linear momenta of electrons in atoms and in solid bodies as revealed by x-ray scattering \textit{Rev. Mod. Phys.} {\bf 5} 1-33

\bibitem{ref1} Curtright T L, Zachos C K 2014 Branched Hamiltonians and supersymmetry \textit{J. Phys. A: Math. and Theor.} {\bf 47} 145201

\bibitem{sw1} Shapere A, Wilczek F 2012 Branched quantization \textit{Phys. Rev. Lett.} {\bf 109} 200402

\bibitem{sw2} Shapere A, Wilczek F 2012 Classical time crystal \textit{Phys. Rev. Lett.} {\bf 109} 160402

\bibitem{w3} Wilczek F 2012 Quantum time crystal \textit{Phys. Rev. Lett.} {\bf 109} 160401

\bibitem{rmc1} Rosu H C, Mancas S C, Chen P 2014  One-parameter families of supersymmetric isospectral potentials \textit{Ann. Phys.} {\bf 343} 87-102

\bibitem{rmc2} Rosu H C, Mancas S C, Chen P 2014  Shifted one-parameter supersymmetric family of quartic asymmetric double-well potentials \textit{Ann. Phys.} {\bf 349} 33-42

\bibitem{miel} Mielnik B 1984 Factorization method and new potentials with the oscillator spectrum \textit{J. Math. Phys.} {\bf 25} 3387-3389

\bibitem{fern} Fern\'andez D 1984 New hydrogen-like potentials \textit{Lett. Math. Phys.} {\bf 8} 337-343

\bibitem{n84} Nieto M M 1984 Relationship between supersymmetry and the inverse scattering method \textit{Phys. Lett. B} {\bf 145} 208-210

\bibitem{r2000} Rosu H C 2000 Multiple parameter structure of Mielnik's isospectrality in unbroken SUSYQM \textit{Int. J. Theor. Phys.} {\bf 39} 105-114


\end{thebibliography}
\end{document}